# Morphology preserving solid-to-solid transformation of minerals mediated by a self-assembled temperature responsive polymer membrane: calcite to hydroxyapatite

*Anders C. S. Jensen[a], Anastasia Brif[b], Boaz Pokroy[b], Mogens Hinge[c] and Henrik Birkedal[a]\**

[a]Department of Chemistry & iNANO, Aarhus University, DK-8000 Aarhus C, Denmark

[b]Department of Materials Science and Engineering & the Russel Berrie Nanotechnology Institute Technion-Israel Institute of Technology, IL-32000 Haifa, Israel

[c]Department of Engineering, Aarhus University, DK-8000 Aarhus C, Denmark

ABSTRACT. Assemblies of nano crystals with complex morphologies have solicited great scientific interest. Controlling the formation of these self-assembled structures has proved difficult. Here we present a novel transformation reaction for transforming a single crystal of minerals into assemblies of nanocrystals of substances with lower solubility without changing the overall morphology. This study focusses on the transformation of rhombohedral calcite crystals to assemblies of hydroxyapatite nanoscrystals (HAP) using poly-(N-isopropyl-acrylamide)-block-

poly-(acrylic-acid) (PNIPAM-b-PAAc) to control the transformation process. This yielded particle assemblies of HAP nanoparticles grossly preserving the original rhombohedral calcite morphology.

## Introduction

Controlling the secondary structure of nanoparticle assemblies has proved a challenge and many different approaches are utilized to achieve this goal. The formation mechanism of crystals has been controlled by precipitating inorganic salts in gel media. In this approach the limiting factor can controlled to be growth kinetics or ion diffusion leading to highly crystalline or dendritic nano crystal assemblies, respectively[1, 2] ENREF_1. The effects of organic additives has also been shown to result in a wide range of morphologies such as spheres[3, 4], hollow spheres[5-10], complex super structures[11-14] and ordered thin films[15]. Disordered transparent materials of densely packed particles have also been made either by using the chemistry inherent to the system[16], assisted by organic additives[17] or controlled and directed by the organic additive[18]. Inorganic precursors can also be used to determine the morphology of the target material. This can be done by using a large single crystal of a less stable polymorph and transforming it into a nano particle assembly of the more stable phase (e.g. octacalcium phosphate to hydroxylapatite (HAP))[19-21]. However, this requires a less stable precursor phase that can transform via solid-to-solid transformation in order to occur.

In this study we present a novel transformation reaction utilizing a thermal responsive block copolymer. The transformation reaction works by dissolving the parent crystals (i.e. calcite) and then repreciptation the target phase (i.e. HAP). The transformation reaction is confined to the surface of the calcite crystals by binding a polymer to the surface and then using the thermal responsivity to form a diffusion limiting membrane. PNIPAM-b-PAAc (Figure 1) was used as the polymer additive. The polymer has an lower critical solubility temperature (LCST) of 32 °C and we have previously shown how this polymer can be used to afford calcite nanocrystals[22]. The PAAc block acts like an anchor binding the polymer to the surface of the calcite crystal. The PNIPAM block act as a membrane, that when in its dehydrated state limits diffusion of ions across the membrane. Reaction with sodium phosphate is herein show to result in the assemblies of HAP nano crystals that maintained the overall morphology of the parent crystal. [BP1]

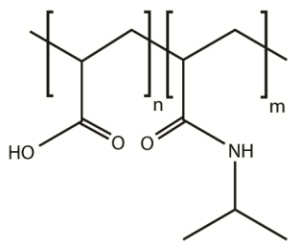

**Figure 1.** PNIPAM-b-PAAc with poly-N-isopropylacrylamide (PNIPAM) on the right and poly acrylic acid (PAAc) on the left.

**Experimental Section**

**Calcite crystal synthesis**. Calcite crystals were synthesized/grown by mixing 20 ml of 0.04 M $CaCl_2 \cdot 2H_2O$ (>99% Sigma Aldrich) with 10 ml of 0.06 M $NaHCO_3$ (>99% Sigma Aldrich) and 0.06 M NaOH (>98% Sigma Aldrich). The reaction was kept at 25 °C for 24 h, using a Julabo ED/F 12 and a water bath with a MIXdrive 15 2Mag magnetic stirrer. The crystals were purified by dialyses for 24 h in DI water and centrifuged. The samples were dried at 60 °C overnight.

**Polymer synthesis.** The polymer synthesis was based on a batch surfactant free emulsion polymerization synthesis developed and reported previously [22],[23]. Briefly, the PNIPAM-b-PAAc was made by dissolving 0.3 g potassium per sulfate (≥99% Sigma-Aldrich) and 6.25 g of NIPAM (97% Sigma Aldrich) in 500 mL of water under $N_2$ gas flow. The solution was heated to 70 °C for 30 min or until stable emulsion was obtained. At this point *0.35* mL AAc (99% Sigma Aldrich) was added obtaining a 5 wt% PAAc weight fraction determined by NMR. Inhibitor in the acrylic acid monomer was prior to addition was removed in an alumina column. The reaction was stirred (350 RPM) for 24 h under reflux and $N_2$ atmosphere, subsequently purified through dialysis against Milli-Q water for 7 day in a spectra/Por® dialysis Membrane MWCO: 3.500 (spectrum laboratories, Inc.).

**Crystal transformation.** 60 mg $CaCO_3$ crystals were added to 20 mL solution of 0.35 mg/mL PNIPAM-b-PAAc at room temperature. The temperature was increase to 60 °C after which 10 mL 36 mM $Na_3PO_4$ (36 mM $NaH_2PO_4$ & 72 mM NaOH) and 12 mM NaOH was added and the reaction was left for 48 h. The samples were washed with DI water and dried at 60 °C. Similar samples were made with a geological sample of Icelandic spar.

### Characterization.

The transformed particles were characterized by SEM and PXRD with Rietveld refinement. For the PXRD a Rigaku Smartlab diffractometer (Rigaku corporation, Tokyo, Japan), with a Cu $K_\alpha$ rotating anode was used in parallel beam focusing mode. The samples were mounted on a flat sample spinner at 30 rpm. A $LaB_6$ standard (NIST) was used to account for the instrumental line broadening. Rietveld refinement was done in Fullprof[24] using spherical harmonics to model the size anisotropy[25]. SEM micrographs were recorded on a Nova nanoSEM 600 (FEI) with an ETD detector. The Icelandic spar sample was cut with a Strata 400 STEM Dual-Beam FIB system, which is a fully digital field emission scanning electron microscope (FEG-SEM) equipped with FIB technology and a flip-stage-STEM assembly. The cutting [HB2]was performed via a gallium beam and examined in a Zeizz UltraSEM (Carl Zeiss AG, Oberkochen, Germany).

### Results and discussion

We made calcite crystals and investigated how they are transformed into hydroxyapatite. The calcite samples were examined by PXRD and SEM prior to the transformation to HAP. PXRD confirmed that the sample was phase pure calcite. SEM showed that the calcite crystals were in the order of 5-10 µm and adopted a rhombohedral morphology as expected (Figure 2).

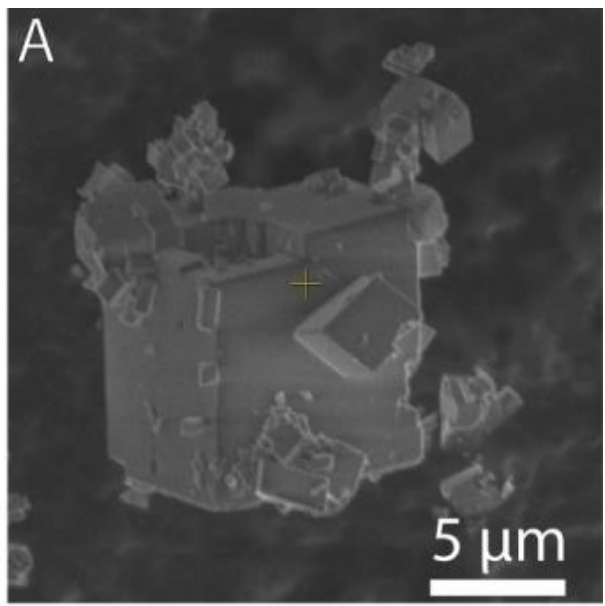

**Figure 2.** SEM micrographs of calcite, showing the regular rhombohedral shape of the calcite crystal.

The calcite crystals were then exposed to the alkaline phosphate solution with and without the polymer. Without the polymer added the calcite crystals slowly dissolved as evident by etch patterns (Figure 3, 25 °C, no additive). As the calcite dissolved, HAP precipitated at random places in the solution at both 25 and 60 °C as seen in Figure 3 (no additive[HB3][BP4]) where the HAP crystals are seen as small aggregates of nanoparticles. At 60 °C the transformation was faster and progressed further towards completion and the SEM micrographs were dominated by the HAP crystals (Figure 3, 60 °C, no additive). When the polymer was added prior to the phosphate solution it binds to the surface of the calcite crystals. At 25 °C (below the LCST) the polymer has little effect and the same etch patterns and random HAP nucleation patterns are seen as without the polymer (Figure 3, 25 °C, PNIPAM-b-PAAc[BP5][HB6]). However, above the LCST drastically different behavior was observed. By adding the polymer at 25 °C it can bind to the surface of the calcite crystals. The PNIPAM block can then be dehydrated by heating above the LCST. In this case the material maintained the overall rhombohedra shape during transformation from single crystal calcite to an assembly of HAP nanocrystals (Figure 3 60°C, PNIPAM-b-PAAc) as confirmed by XRD below. We suggest that this behavior can be explained by the formation of a polymer membrane around the calcite crystals that forms a barrier for both dissolution and crystallization and thus only affords hydroxyapatite formation at the calcite crystal surface.

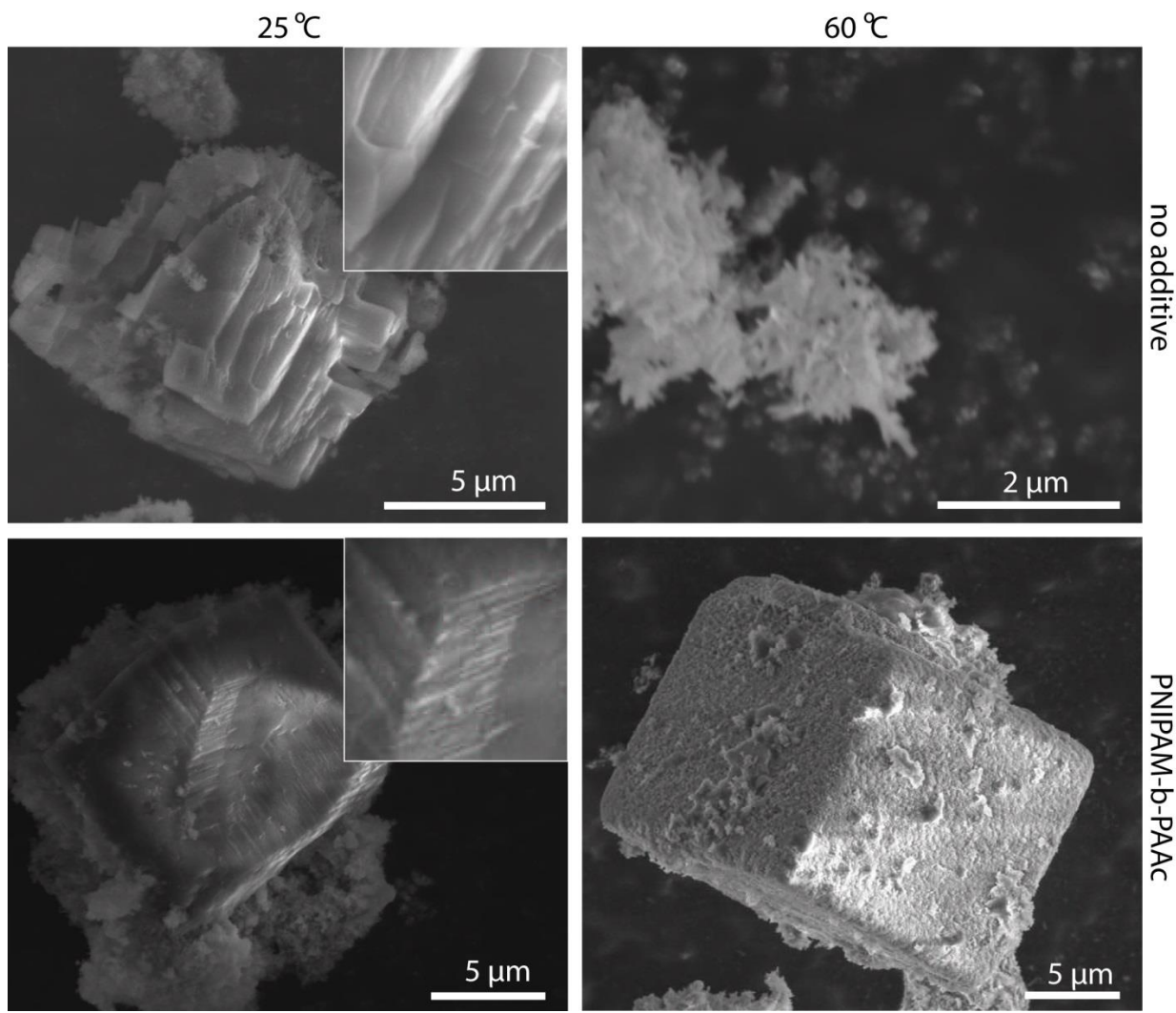

**Figure 3.** SEM micrographs of calcite crystals after 48 h in a 12 mM Na$_3$PO$_4$, 4 mM NaOH

solution without (top) and with (bottom) PNIPAM-b-PAAc block copolymer additive added.

The PXRD data of the transformed samples could be explained by a mixture of calcite and

hydroxyapatite (Figure 4A). From the analysis of the weight fractions of the crystalline materials

in the sample it is clear that the transformation without polymer present is much faster at 60 °C

then at 25 °C as expected (Figure 4B). However, with the polymer added the degree of transformation is only slightly faster at 60°C than at 25 °C (Figure 4b) and the strong acceleration seen in the polymer-free system is not observed. This is consistent with the limited diffusion of ions due to the hydrophobic PNIPAM block at this temperature. The rate of transformation can be increased by increasing the phosphate concentration (Figure 4C). This is reasonable as a higher concentration will increase the ionic gradient across the polymer membrane. These samples also yielded rhombohedral particles.

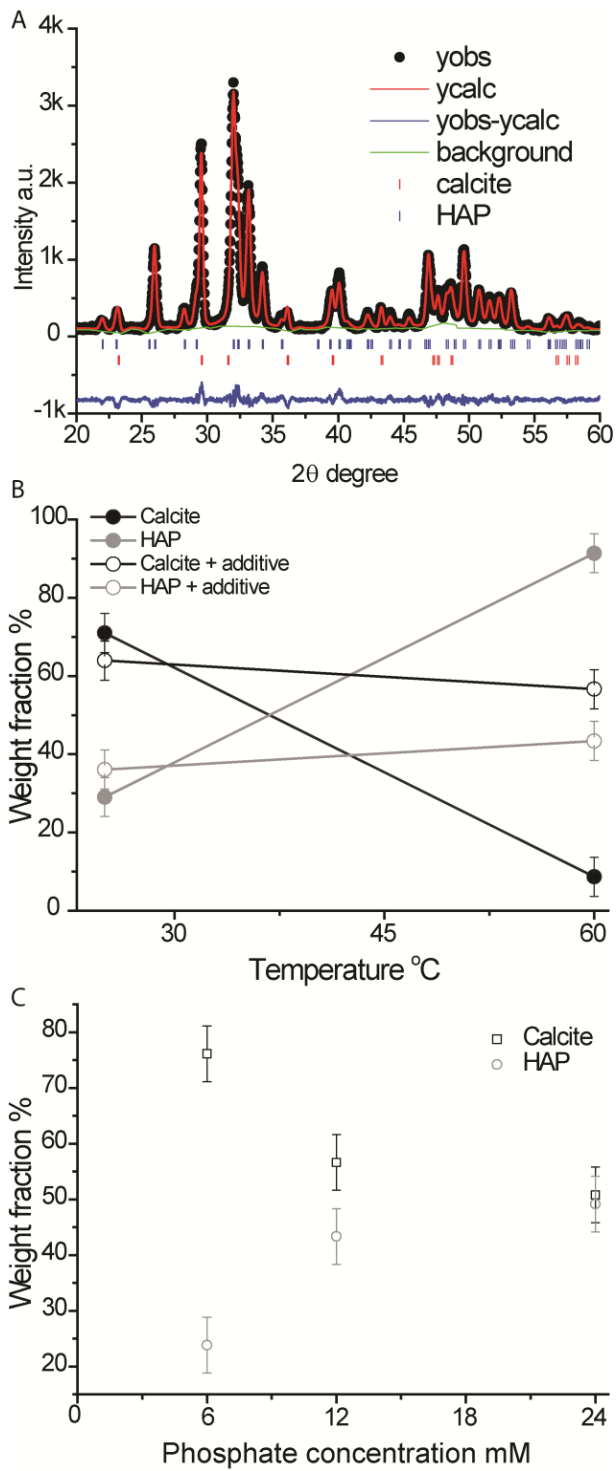

**Figure 4.** Rietveld refinement of XRD data. (A) Example diffractogram of the transformed calcite/HAP sample. (B) Refined weight fraction of the calcite/HAP samples transformed with (open symbols) and without ([HB7]closed symbols)[BP8] the polymer additive above and below the LCST (32 °C) (B). Refined weight fraction of the calcite/HAP samples transformed above the LCST at various phosphate concentrations (C).

To get a better image of the calcite/HAP interface a geological sample of Icelandic spar (geological calcite) was crushed and transformed as described above. A corner of the partially transformed sample was then cut and imaged given in Figure 5. A sharp interphase between the two phases was found. Some areas showed a well-defined interphase with no observable transition between the two phases (Figure 5b) while other areas showed a more gradual transition (Figure 5c). This supports the reaction mechanism of the polymer membrane that limits the reaction to the surface of the calcite crystal. While this result is similar to what is expected of a solid to solid transformation, it is in this case mediated by the solvent (dissolution/recrystallization). But the polymer membrane localizes this transformation zone to the near surface region around the dissolving calcite crystal.

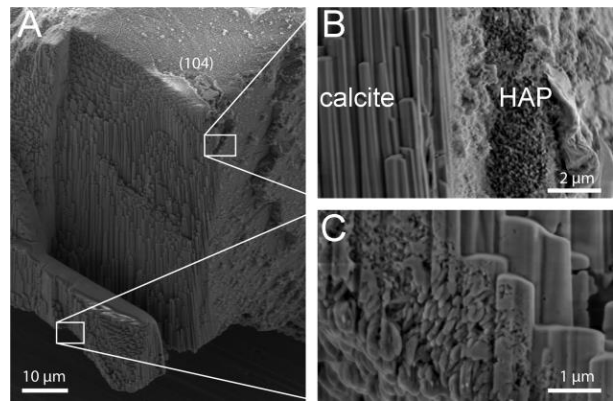

**Figure 5.** SEM micrographs of a geological calcite sample post transformation showing an overview of a corner exposed by FIB, the column structure on the exposed surface is damage from the FIB (a). Zoom of a calcite /HAP  interface with a sharp (b) and gradual interface (c).

The reaction is summarized in Figure 6. As shown by SEM the transformation yields random nucleation of HAP nanoparticles if the transformation is performed without the polymer additive (6A) or with the polymer below the LCST (6C). The morphology is only preserved by first binding the polymer to the surface forming a membrane (6B), condensation of the polymer membrane by increasing the temperature above the LCST (6D) and then initiating the transformation by adding the phosphate solution (6E).

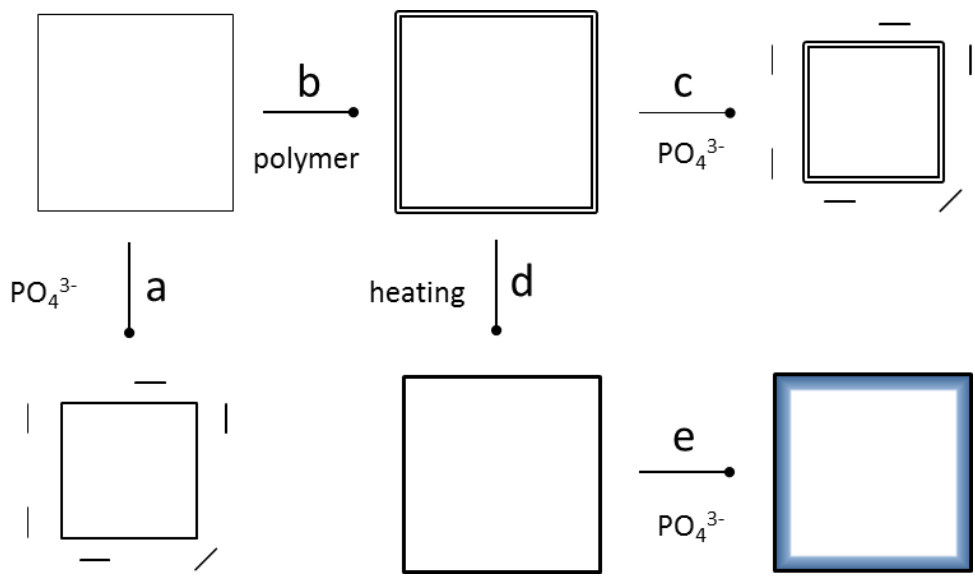

**Figure 6.** S[MH9]chematic representation of the reaction mechanism. Transforming the calcite without the polymer leads to dissolution of the calcite crystal (white square) and random nucleation of HAP nano particles (black lines) (a). Adding the polymer below the LCST allows for the carboxylate groups of the PAAc block to bind to the calcite surface (b). Transforming the calcite crystal below the LCST yields the same result as in a (c). Increasing the temperature above the LCST expels the solvent from the PNIPAM block (d). Now the transformation is confined by the limited diffusion of the PNIPAM block and the overall morphology is conserved throughout the reaction (e)

This type of reaction mechanism is potentially universal as there are only two prerequisites: 1) that the parent material is more soluble then the target material. 2) That the binding block (i.e. PAAc) can bind strong enough to the parent crystal to form the membrane. This opens up to a whole series of chemical transformations into a wide range of morphologies (e.g. aragonite, vaterite or ACC transformed to HAP). Such interfacial crystallization regimes have been shown to give rise to astounding morphologies e.g. at liquid-liquid interfaces in chemical garden systems[26, 27].

**Conclusion.**

We have successfully transformed calcite crystals to assemblies of HAP nano crystals while preserving the macro-morphology of the calcite crystal. A reaction mechanism has been suggested indicating that this type of reaction can have a great potential for using polymeric scaffolds to control the morphology of a material.

ASSOCIATED CONTENT

**Supporting Information**.

AUTHOR INFORMATION

**Corresponding Author**

Prof. Dr. Henrik Birkedal

Aarhus University

Department of Chemistry and iNANO

Email:hbirkedal@chem.au.dk

**Author Contributions**

The manuscript was written through contributions of all authors. All authors have given approval to the final version of the manuscript.

**Funding Sources**

**Notes**

ACKNOWLEDGMENT

We thank Ms. Maria Koifman and Dr. Alex Berner for help with the FIB and SEM analysis and Mr. Thomas Feldberg for assisting with the synthesis of PNIPAM-b-PAAc.

BP acknowledges support from the European Research Council under the European Union's Seventh Framework Program (FP/2007–2013)/ERC Grant Agreement n° [336077].

ABBREVIATIONS

REFERENCES

(Word Style "TF_References_Section"). References are placed at the end of the manuscript. Authors are responsible for the accuracy and completeness of all references. Examples of the recommended format for the various reference types can be found at

http://pubs.acs.org/page/4authors/index.html. Detailed information on reference style can be found in *The ACS Style Guide,* available from Oxford Press.

BRIEFS (Word Style "BH_Briefs"). If you are submitting your paper to a journal that requires a brief, provide a one-sentence synopsis for inclusion in the Table of Contents.

SYNOPSIS (Word Style "SN_Synopsis_TOC"). If you are submitting your paper to a journal that requires a synopsis, see the journal's Instructions for Authors for details.